# Convective heat transfer to Sisko fluid over a rotating disk

*Asif Munir[1] and Masood Khan*

*Department of Mathematics, Quaid-i-Azam University, Islamabad 44000, Pakistan*

**Abstract**: This article deals with study of the steady flow and heat transfer characteristics of Sisko fluid over a rotating infinite disk. The flow and heat transfer aspects are thoroughly investigated encompassing highly shear thinning/thickening Sisko fluids. The modeled boundary layer equations are reduced to a system of nonlinear ordinary differential equations using the appropriate transformation. The resulting equations are then solved numerically by shooting method in the domain $[0,\infty)$. The numerical data for the velocity and temperature fields are graphically sketched and effects of the relevant parameters are discussed in detail. In addition, the velocity gradients at the disk surface and the local Nusselt number for different values of the pertaining parameters are given in tabulated form. Further, the flow and temperature fields of power-law and Newtonian fluids are also compared with those Sisko fluid. Moreover, a comparison with previously published work, as a special case of the problem, has been provided and the results are found to be in excellent agreement.

## 1. Introduction

Rotating disk systems are widely used in engineering applications, such as power engineering, chemical, oil and food processing industry, medical equipment and viscometers *etc.* [1]. Large rotating flow systems are present in the atmosphere, in oceans and around the famous Great Red spot of the planet Jupiter [2]. In 1920, Von Karman [3] initiated the work on these problems by considering the flow induced by infinite rotating disk, where the fluid far from the disk was at rest. This endeavor opened the way for subsequent explorations on rotating disks. The corresponding heat transfer problems with different aspects were investigated by several researchers like Millsaps and Pohlhausen [4], Sparrow and Gregg [5], Riley [6], Awad [7] *etc*. [4-7]. Turkyilmazoglu [8, 9] studied in depth the effects of radial electric field on flow and heat transfer due to the rotating disk. Later on, the same author [10,11] analyzed the effects of stretching and shrinking rotating disks on flow and heat transfer in electrically conducting fluids.

---
[1]Corresponding author: asifmunir1000@yahoo.com

All of the aforementioned studies lie in the domain of Newtonian fluids. There is a vast utilization of non-Newtonian fluids in industrial sector such as pharmaceutical, polymer, personal care products and so forth. Various researchers explored the flow and heat transfer to non-Newtonian fluids due to a rotating infinite disk, including Mitschka and Ulbrecht [12]. They devoted their attention to the rotating disk flow of a power-law fluid and obtained the similar solutions. Gorla [13] addressed the heat transfer in boundary layer flow of power-law fluid past a rotating disk with a step discontinuity in the surface temperature. In two separate studies, Andersson *et al.* [14,15] analyzed the effects of magnetic field on the flow of an electrically conducting power-law fluid in the vicinity of a constantly rotating infinite disk. Attia [16] explored the steady flow of an incompressible viscous non-Newtonian fluid above an infinite rotating porous disk in a porous medium with heat transfer. Ming *et al.* [17] investigated the steady flow and heat transfer of the power-law fluid over a rotating disk. An addition of Newtonian contribution to the power-law model leads to the Sisko fluid model. Khan and Shahzad [18,19] comprehensively analyzed the flow of Sisko fluid in different flow configurations. A through study regarding flow and convective heat transfer is dealt with by Munir *et al.* [20].

Keeping in view the above mentioned literature survey, it safe to conclude that still no study has been reported regarding flow and heat transfer to Sisko fluids above an infinite rotating disk and this article intends to focus this problem thoroughly to fill this gap in the relevant literature.

## 2.0   Mathematical model and formulation

## 2.1   Boundary layer equations

To develop the boundary layer equations for Sisko fluid over an infinite rotating disk, we take the cylindrical polar coordinate system $(r, \phi, z)$. For a three dimensional axisymmetric flow the velocity, stress and temperature fields have the form

$$\mathbf{V} = \big[ u(r, \phi, z), v(r, \phi, z), w(r, \phi, z) \big], \ \mathbf{S} = \mathbf{S}(r, \phi, z), \ T = T(r, \phi, z), \tag{1}$$

where $u, v$ and $w$ are the radial, azimuthal and axial components of the velocity vector, respectively.

The continuity, momentum and thermal energy equations for the steady flow of an incompressible Sisko fluid simplify to

$$\frac{\partial u}{\partial r} + \frac{u}{r} + \frac{\partial w}{\partial z} = 0, \tag{2}$$

$$\rho \left( u \frac{\partial u}{\partial r} - \frac{v^2}{r} + w \frac{\partial u}{\partial z} \right) = -\frac{\partial p}{\partial r} + \frac{1}{r} \frac{\partial (r s_{rr})}{\partial r} + \frac{\partial (s_{zr})}{\partial z} - \frac{s_{\phi\phi}}{r}, \tag{3}$$

$$\rho \left( u \frac{\partial v}{\partial r} + \frac{uv}{r} + w \frac{\partial v}{\partial z} \right) = -\frac{\partial p}{\partial \phi} + \frac{1}{r^2} \frac{\partial \left( r^2 s_{r\phi} \right)}{\partial r} + \frac{\partial (s_{z\phi})}{\partial z} + \frac{s_{\phi r} - s_{r\phi}}{r}, \tag{4}$$

$$\rho\left(u\frac{\partial w}{\partial r}+w\frac{\partial w}{\partial z}\right)=-\frac{\partial p}{\partial z}+\frac{1}{r}\frac{\partial(rs_{rz})}{\partial r}+\frac{\partial(s_{zz})}{\partial z},\tag{5}$$

$$\rho c_p\left(u\frac{\partial T}{\partial r}+w\frac{\partial T}{\partial z}\right)=\kappa\frac{\partial^2 T}{\partial z^2}+\frac{1}{r}\frac{\partial T}{\partial r}+\frac{\partial^2 T}{\partial r^2},\tag{6}$$

$$\left.\begin{array}{l}s_{\phi r}=s_{r\phi}=\left(a+b\left|\frac{1}{2}tr\mathbf{A}_1^2\right|^{\frac{n-1}{2}}\right)\left(\frac{\partial v}{\partial r}-\frac{v}{r}\right),\; s_{rr}=\left(a+b\left|\frac{1}{2}tr\mathbf{A}_1^2\right|^{\frac{n-1}{2}}\right)\left(2\frac{\partial u}{\partial r}\right),\\[3mm]
s_{zr}=s_{rz}=\left(a+b\left|\frac{1}{2}tr\mathbf{A}_1^2\right|^{\frac{n-1}{2}}\right)\left(\frac{\partial u}{\partial z}+\frac{\partial w}{\partial r}\right),\; s_{\phi\phi}=\left(a+b\left|\frac{1}{2}tr\mathbf{A}_1^2\right|^{\frac{n-1}{2}}\right)\left(2\frac{u}{r}\right),\\[3mm]
s_{z\phi}=s_{\phi z}=\left(a+b\left|\frac{1}{2}tr\mathbf{A}_1^2\right|^{\frac{n-1}{2}}\right)\left(\frac{\partial v}{\partial z}\right),\qquad s_{zz}=\left(a+b\left|\frac{1}{2}tr\mathbf{A}_1^2\right|^{\frac{n-1}{2}}\right)\left(2\frac{\partial w}{\partial z}\right),\end{array}\right\}\tag{7}$$

$$\left|\frac{1}{2}tr\mathbf{A}_1^2\right|^{\frac{n-1}{2}}=\left|2\left(\frac{\partial u}{\partial r}\right)^2+\left(\frac{\partial v}{\partial r}-\frac{v}{r}\right)^2+\left(\frac{\partial u}{\partial z}-\frac{\partial w}{\partial r}\right)^2+\left(\frac{\partial v}{\partial z}\right)^2+2\left(\frac{u}{r}\right)^2+2\left(\frac{\partial w}{\partial z}\right)^2\right|^{\frac{n-1}{2}}.\tag{8}$$

By assuming the symmetry about the axis of rotation $\left(\frac{\partial}{\partial\varphi}(.)=0\right)$ [14] and introducing non-dimensional variables to deduce the boundary layer equations as

$$\tilde{u}=\frac{u}{R\Omega},\;\tilde{v}=\frac{v}{R\Omega},\;\tilde{w}=\frac{w}{\Omega\delta},\;\tilde{r}=\frac{r}{R},\;\tilde{z}=\frac{z}{\delta},\;\tilde{T}=\frac{T-T_\infty}{T_\infty},\;\tilde{p}=\frac{p}{\rho(R\Omega)^2}.\tag{9}$$

Accordingly non-dimensional form of the above equations can be written as

$$\frac{\partial\tilde{u}}{\partial\tilde{r}}+\frac{\tilde{u}}{\tilde{r}}+\frac{\partial\tilde{w}}{\partial\tilde{z}}=0,\tag{10}$$

$$\tilde{u}\frac{\partial\tilde{u}}{\partial\tilde{r}}+\tilde{w}\frac{\partial\tilde{u}}{\partial\tilde{z}}-\frac{\tilde{v}^2}{\tilde{r}}=-\frac{\partial\tilde{p}}{\partial\tilde{r}}+\frac{a/\rho}{\Omega R^2}\frac{\partial^2\tilde{u}}{\partial\tilde{r}^2}+\frac{a/\rho}{\Omega R^2}\left(\frac{R}{\delta}\right)^2\frac{\partial^2\tilde{u}}{\partial\tilde{z}^2}-\frac{a/\rho}{\Omega R^2}\frac{\tilde{u}}{\tilde{r}^2}+2\frac{a/\rho}{\Omega R^2}\frac{\partial\tilde{u}}{\partial\tilde{r}}$$

$$+2\frac{b/\rho}{(\Omega R)^{2-n}}R^n\left(\frac{R}{\delta}\right)^{n+1}\frac{1}{(R/\delta)^2}\left[\frac{\partial}{\partial\tilde{r}}\left(\left(\frac{\partial\tilde{u}}{\partial\tilde{r}}\right)\tilde{J}\right)+\frac{1}{\tilde{r}}\left(\frac{\partial\tilde{u}}{\partial\tilde{r}}\right)\tilde{J}-\frac{\tilde{u}}{\tilde{r}^2}\tilde{J}\right]$$

$$+\frac{b/\rho}{(\Omega R)^{2-n}}R^n\left(\frac{R}{\delta}\right)^{n+1}\frac{\partial}{\partial\tilde{z}}\left[\left(\frac{\partial\tilde{u}}{\partial\tilde{z}}+\frac{1}{(R/\delta)^2}\frac{\partial\tilde{w}}{\partial\tilde{r}}\right)\tilde{J}\right],\tag{11}$$

$$\tilde{u}\frac{\partial\tilde{v}}{\partial\tilde{r}}+\tilde{w}\frac{\partial\tilde{v}}{\partial\tilde{z}}+\frac{\tilde{u}\tilde{v}}{\tilde{r}}=\frac{a/\rho}{\Omega R^2}\frac{\partial^2\tilde{v}}{\partial\tilde{r}^2}+\frac{a/\rho}{\Omega R^2}\left(\frac{R}{\delta}\right)^2\frac{\partial^2\tilde{v}}{\partial\tilde{z}^2}-\frac{a/\rho}{\Omega R^2}\frac{\tilde{v}}{\tilde{r}^2}+2\frac{a/\rho}{\Omega R^2}\frac{1}{\tilde{r}}\frac{\partial\tilde{v}}{\partial\tilde{r}}$$

$$+2\frac{b/\rho}{(\Omega R)^{2-n}R^n}\left(\frac{R}{\delta}\right)^{n+1}\frac{1}{(R/\delta)^2}\left[\frac{\partial}{\partial\tilde{r}}\left(\left(\frac{\partial\tilde{v}}{\partial\tilde{r}}-\frac{\tilde{v}}{\tilde{r}}\right)\tilde{J}\right)+\frac{1}{\tilde{r}}\left(\frac{\partial\tilde{v}}{\partial\tilde{r}}-\frac{\tilde{v}}{\tilde{r}}\right)\tilde{J}\right]$$

$$+\frac{b/\rho}{(\Omega R)^{2-n}R^n}\left(\frac{R}{\delta}\right)^{n+1}\frac{\partial}{\partial\tilde{z}}\left[\left(\frac{\partial\tilde{v}}{\partial\tilde{z}}\right)\tilde{J}\right],\tag{12}$$

$$\frac{1}{(R/\delta)^2}\left(\tilde{u}\frac{\partial\tilde{w}}{\partial\tilde{r}}+\tilde{w}\frac{\partial\tilde{w}}{\partial\tilde{z}}\right)=-\frac{\partial\tilde{p}}{\partial\tilde{z}}+\frac{a/\rho}{\Omega R^2}\frac{\partial^2\tilde{w}}{\partial\tilde{z}^2}+\frac{a/\rho}{\Omega R^2}\left(\frac{1}{R/\delta}\right)^2\frac{\partial^2\tilde{w}}{\partial\tilde{r}^2}+\frac{a/\rho}{\Omega R^2}\frac{1}{\tilde{r}}\frac{\partial\tilde{u}}{\partial\tilde{z}}$$

$$+\frac{a/\rho}{\Omega R^2}\frac{1}{(R/\delta)^2}\frac{1}{\tilde{r}}\frac{\partial\tilde{w}}{\partial\tilde{r}}+\frac{b/\rho}{(\Omega R)^{2-n}R^n}\left(\frac{R}{\delta}\right)^{n+1}\frac{1}{(R/\delta)^2}\left[\frac{\partial}{\partial\tilde{r}}\left(\left(\frac{\partial\tilde{u}}{\partial\tilde{z}}+\frac{1}{(R/\delta)^2}\frac{\partial\tilde{w}}{\partial\tilde{r}}\right)\tilde{J}\right)\right.$$

$$\left.+2\frac{\partial}{\partial\tilde{z}}\left(\left(\frac{\partial\tilde{w}}{\partial\tilde{z}}\right)\tilde{J}\right)+\left(\frac{\partial\tilde{u}}{\partial\tilde{z}}+\frac{1}{(R/\delta)^2}\frac{\partial\tilde{w}}{\partial\tilde{r}}\right)\tilde{J}\right],\tag{13}$$

where

$$\tilde{J}=\left(\frac{R\Omega}{\delta}\right)^{n-1}\left|\frac{1}{(R/\delta)^2}\left[2\left(\frac{\partial\tilde{u}}{\partial\tilde{r}}\right)^2+\left(\frac{\partial\tilde{v}}{\partial\tilde{r}}\right)^2+\frac{1}{(R/\delta)^2}\left(\frac{\partial\tilde{w}}{\partial\tilde{r}}\right)^2+2\left(\frac{\partial\tilde{w}}{\partial\tilde{z}}\right)^2\right.\right.$$

$$\left.\left.+2\left(\frac{\partial\tilde{w}}{\partial\tilde{r}}\right)\left(\frac{\partial\tilde{u}}{\partial\tilde{z}}\right)-2\frac{\tilde{v}}{\tilde{r}}\frac{\partial\tilde{v}}{\partial\tilde{r}}+\left(\frac{\tilde{v}}{\tilde{r}}\right)^2+\left(\frac{\tilde{u}}{\tilde{r}}\right)^2\right]+\left(\frac{\partial\tilde{u}}{\partial\tilde{z}}\right)^2+\left(\frac{\partial\tilde{v}}{\partial\tilde{z}}\right)^2\right|^{\frac{n-1}{2}}\tag{14}$$

$$\tilde{u}\frac{\partial\tilde{T}}{\partial\tilde{r}}+\tilde{w}\frac{\partial\tilde{T}}{\partial\tilde{z}}=\frac{1}{\Omega R^2}\left(\frac{\kappa}{\rho c_p}\right)\left[\frac{\partial^2\tilde{T}}{\partial\tilde{z}^2}+\frac{1}{\tilde{r}}\frac{\partial\tilde{T}}{\partial\tilde{r}}+\left(\frac{R}{\delta}\right)^2\frac{\partial^2\tilde{T}}{\partial\tilde{r}^2}\right].\tag{15}$$

Within the boundary layer the inertial and viscous terms have the same order of magnitude and these terms give us

$$\frac{a/\rho}{\Omega R^2}\left(\frac{R}{\delta}\right)^2=O(1),\ \left(\frac{\delta}{R}\right)^2=O\left(\frac{1}{\mathrm{Re}_a}\right)\text{ and }\frac{b/\rho}{R^n(\Omega R)^{2-n}}\left(\frac{R}{\delta}\right)^{n+1}=O(1),\tag{16}$$

where $\mathrm{Re}_a=\frac{\Omega R^2}{a/\rho}$. In the limit $\mathrm{Re}_a\to\infty$, Eqs.(10) to (15) asymptotically become

$$\tilde{u}\frac{\partial\tilde{u}}{\partial\tilde{r}}+\tilde{w}\frac{\partial\tilde{u}}{\partial\tilde{z}}-\frac{\tilde{v}^2}{\tilde{r}}=-\frac{\partial\tilde{p}}{\partial\tilde{r}}+\frac{\partial^2\tilde{u}}{\partial\tilde{z}^2}+\frac{\partial}{\partial\tilde{z}}\left[\left(\frac{\partial\tilde{u}}{\partial\tilde{z}}\right)\left|\left(\frac{\partial\tilde{u}}{\partial\tilde{z}}\right)^2+\left(\frac{\partial\tilde{v}}{\partial\tilde{z}}\right)^2\right|^{\frac{n-1}{2}}\right],\tag{17}$$

$$\tilde{u}\frac{\partial \tilde{v}}{\partial \tilde{r}} + \tilde{w}\frac{\partial \tilde{v}}{\partial \tilde{z}} + \frac{\tilde{u}\tilde{v}}{\tilde{r}} = \frac{\partial^2 \tilde{v}}{\partial \tilde{z}^2} + \frac{\partial}{\partial \tilde{z}}\left[\left(\frac{\partial \tilde{v}}{\partial \tilde{z}}\right)\left|\left(\frac{\partial \tilde{u}}{\partial \tilde{z}}\right)^2 + \left(\frac{\partial \tilde{v}}{\partial \tilde{z}}\right)^2\right|^{\frac{n-1}{2}}\right], \tag{18}$$

$$0 = -\frac{\partial \tilde{p}}{\partial \tilde{z}}, \tag{19}$$

and

$$\tilde{u}\frac{\partial \tilde{T}}{\partial \tilde{r}} + \tilde{w}\frac{\partial \tilde{T}}{\partial \tilde{z}} = \frac{\kappa}{\rho c_p}\frac{\partial^2 \tilde{T}}{\partial \tilde{z}^2}. \tag{20}$$

The above equations in dimensional form turn out to be the following:

$$\frac{\partial u}{\partial r} + \frac{u}{r} + \frac{\partial w}{\partial z} = 0, \tag{21}$$

$$u\frac{\partial u}{\partial r} + w\frac{\partial u}{\partial z} - \frac{v^2}{r} = -\frac{\partial p}{\partial r} + a\frac{\partial^2 u}{\partial z^2} + b\frac{\partial}{\partial z}\left[\left(\frac{\partial u}{\partial z}\right)\left|\left(\frac{\partial u}{\partial z}\right)^2 + \left(\frac{\partial v}{\partial z}\right)^2\right|^{\frac{n-1}{2}}\right], \tag{22}$$

$$u\frac{\partial v}{\partial r} + w\frac{\partial v}{\partial z} + \frac{uv}{r} = a\frac{\partial^2 v}{\partial z^2} + b\frac{\partial}{\partial z}\left[\left(\frac{\partial v}{\partial z}\right)\left|\left(\frac{\partial u}{\partial z}\right)^2 + \left(\frac{\partial v}{\partial z}\right)^2\right|^{\frac{n-1}{2}}\right], \tag{23}$$

$$0 = -\frac{\partial p}{\partial z}, \tag{24}$$

$$u\frac{\partial T}{\partial r} + w\frac{\partial T}{\partial z} = \frac{\kappa}{\rho c_p}\frac{\partial^2 T}{\partial z^2}. \tag{25}$$

## 2.2 Governing problem

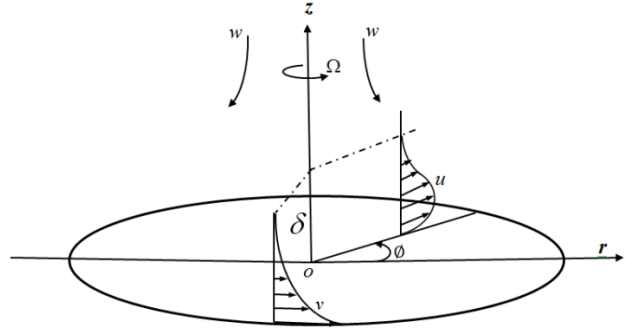

***Figure 1:*** Physical model and coordinate system.

We will consider the steady axisymmetric flow of a Sisko fluid over an infinite disk rotating with a sufficiently high constant angular velocity $\Omega$, so that effects of gravitational force on momentum transfer are rather low. The disk coincides with the plane $z = 0$ and the motion of the fluid is confined in the half space $z > 0$. In a cylindrical coordinates system, the disk rotates around its axis of symmetry coinciding with the $z$-axis. The disk is kept at a constant temperature $T_w$ and the fluid outside the boundary layer is maintained at a uniform temperature $T_\infty$. The physical model in cylindrical polar coordinate system is shown through figure 1 while the governing boundary layer equations take the form

$$u\frac{\partial u}{\partial r} + w\frac{\partial u}{\partial z} - \frac{v^2}{r} = -\frac{\partial p}{\partial r} + a\frac{\partial^2 u}{\partial z^2} + b\frac{\partial}{\partial z}\left[\left(\frac{\partial u}{\partial z}\right)\left\{\left(\frac{\partial u}{\partial z}\right)^2 + \left(\frac{\partial v}{\partial z}\right)^2\right\}^{\frac{n-1}{2}}\right],$$ (26)

$$u\frac{\partial v}{\partial r} + w\frac{\partial v}{\partial z} + \frac{uv}{r} = a\frac{\partial^2 v}{\partial z^2} + b\frac{\partial}{\partial z}\left[\left(\frac{\partial v}{\partial z}\right)\left\{\left(\frac{\partial u}{\partial z}\right)^2 + \left(\frac{\partial v}{\partial z}\right)^2\right\}^{\frac{n-1}{2}}\right],$$ (27)

$$u\frac{\partial T}{\partial r} + w\frac{\partial T}{\partial z} = \frac{\kappa}{\rho c_p}\frac{\partial^2 T}{\partial z^2}.$$ (28)

The boundary conditions relating to the flow and temperature fields are

$$u = w = 0, \; v = \Omega r \text{ and } T = T_w \text{ at } z = 0,$$ (29)

$$u, \; v \to 0 \text{ and } T \to T_\infty \text{ as } z \to +\infty.$$ (30)

Now introducing the Von Ka′rma′n type of the velocity field and the dimensionless variables defined by

$$\eta = z \left( \frac{\Omega^{2-n}}{b/\rho} \right)^{1/(n+1)} r^{(1-n)/(1+n)}, \ u = r\Omega f(\eta), \ v = r\Omega g(\eta),$$

$$w = z \left( \frac{\Omega^{1-2n}}{b/\rho} \right)^{-1/(n+1)} r^{(n-1)/(n+1)} h(\eta), \text{ and } \theta(\eta) = \frac{T - T_\infty}{T_w - T_\infty}.$$

(31)

Utilizing the variables defined in Eq. (31), the boundary layer Eqs. (26) to (30), in the absence of radial pressure gradient [14] can be written as

$$h' = -2f - \frac{1-n}{1+n} \eta f', \tag{32}$$

$$Af'' + \left[ f' \left( f'^2 + g'^2 \right)^{\frac{n-1}{2}} \right]' = f^2 - g^2 + \left[ \eta \left( \frac{1-n}{1+n} \right) f + h \right] f', \tag{33}$$

$$Ag'' + \left[ g' \left( f'^2 + g'^2 \right)^{\frac{n-1}{2}} \right]' = 2fg + \left[ \eta \left( \frac{1-n}{1+n} \right) f + h \right] g', \tag{34}$$

$$\theta'' \left( f'^2 + g'^2 \right)^{\frac{n-1}{2}} + (n-1)\theta' \left( f'f'' + g'g'' \right) \left( f'^2 + g'^2 \right)^{\frac{n-3}{2}} = \Pr \left( \frac{1-n}{1+n} \eta f + h \right) \theta'. \tag{35}$$

The transformed boundary conditions are

$$f = h = 0, \ g = 1 \text{ at } \eta = 0, \tag{36}$$

$$f = 0, \ g = 0 \text{ as } \eta \to \infty. \tag{37}$$

## 3. Numerical solution of the problem

The exact and/or closed form solution of nonlinear ordinary differential equation is not always straightforward to obtain. So, one has to recourse to the numerical solution of transformed ordinary differential equations, with the associated boundary conditions. Eqs. (32) to (35), alongside the boundary conditions (36) and (37) constitute a two-point boundary value problem and are solved by employing shooting method using adaptive Runge Kutta quadrature with Newton's method in the domain $[0, \infty)$. These equations are written as an equivalent first order system in $\eta$ as follows:

$$f' = p, \tag{38}$$

$$p' = q, \tag{39}$$

$$\theta' = t, \tag{40}$$

$$N_1 = (n-1)pq\left(p^2 + q^2\right)^{\frac{n-3}{2}}\left[2fg + \left\{h + \eta\left(\frac{1-n}{1+n}\right)f\right\}q\right]$$

$$+\left(f^2 - g^2\right)\left[A + \left(p^2 + q^2\right)^{\frac{n-1}{2}} + (n-1)q^2\left(p^2 + q^2\right)^{\frac{n-3}{2}}\right]$$

$$+p\left[\eta\left(\frac{1-n}{1+n}\right)f + h\right]\left[A + \left(p^2 + q^2\right)^{\frac{n-1}{2}} + (n-1)\left(p^2 + q^2\right)^{\frac{n-3}{2}}\right], \tag{41}$$

$$N_2 = (n-1)pq\left(p^2 + q^2\right)^{\frac{n-3}{2}}\left[\left(f^2 - g^2\right) + \left\{h + \eta\left(\frac{1-n}{1+n}\right)f\right\}p\right]$$

$$+2fg\left[A + \left(p^2 + q^2\right)^{\frac{n-1}{2}} + (n-1)p^2\left(p^2 + q^2\right)^{\frac{n-3}{2}}\right]$$

$$+q\left[\eta\left(\frac{1-n}{1+n}\right)f + h\right]\left[A + \left(p^2 + q^2\right)^{\frac{n-1}{2}} + (n-1)p^2\left(p^2 + q^2\right)^{\frac{n-3}{2}}\right], \tag{42}$$

$$D = \left[A + \left(p^2 + q^2\right)^{\frac{n-1}{2}}\right]^2 - \left[(n-1)pq\left(p^2 + q^2\right)^{\frac{n-3}{2}}\right]^2$$

$$+(n-1)^2\left(pq\right)^2\left(p^2 + q^2\right)^{n-3} + \left[A + \left(p^2 + q^2\right)^{\frac{n-1}{2}}\right]\left[(n-1)\left(p^2 + q^2\right)^{\frac{n-1}{2}}\right], \tag{43}$$

$$p' = \frac{N_1}{D}, \qquad (44)$$

$$q' = \frac{N_2}{D}, \qquad (45)$$

$$t' = \left(p^2 + q^2\right)^{\frac{1-n}{2}} \left[ (1-n)t\left(p^2 + q^2\right)^{\frac{n-3}{2}}\left(pp' + qq'\right) + \Pr\left\{\left(\frac{1-n}{1+n}\right)\eta f + h\right\}t\right]. \qquad (46)$$

The corresponding boundary conditions are

$$f(0) = 0, \; g(0) = 1, \; h(0) = 0 \text{ and } \theta(0) = 1, \qquad (47)$$

$$f \to 0, \; g \to 0, \; \theta \to 0 \text{ as } \eta \to \infty. \qquad (48)$$

## 4.  Validation of numerical results

Checking the credibility of our numerical predictions is strongly recommended in any case. The present results are compared with the previously published literature as a special case of our problem. Tables 1(a-c) present the gradient of radial and azimuthal velocity at the wall, as computed in present special case, predicted by Anderson *et al.* [14] and obtained by Mitschka and Ulbrecht (M&U) [12]. A good agreement in these results enhances the veracity of our results. Further, Table 1(c) shows the axial velocity field far from the rotating disk as depicted by our computations and demonstrated in [14]. An encouraging comparison is observed in this case also.

## 5.  Numerical results and discussion

To get the physical insight of the problem, the coupled nonlinear ordinary differential equations (32) to (35) subject to boundary conditions (36) and (37) have been solved numerically. The effects of pertinent non-dimensional parameters influencing the flow and temperature fields to Sisko fluid over a rotating infinite disk with their respective ranges focused are: the power-law index ($0.3 \leq n \leq 1.9$), the material parameter of Sisko fluid ($0.5 \leq A \leq 2$), and Prandtl number ($1 \leq \Pr \leq 4$).

Figures (2) to (12) display the numerical data generated for the flow and temperature fields. The flow field includes the radial $f(\eta)$, azimuthal $g(\eta)$ and axial $h(\eta)$ components of velocity field. The entire velocity field for the shear thinning ($n < 1$) and shear thickening ($n > 1$) Sisko fluids are presented in figures (2) to (4). On inspection of figures 2 (a,b) it is obvious that the radial velocity decreases as the value of the power-law index incremented progressively. A peak near the disk is observed due to the presence of centrifugal force; a slight variation in peak is observed for shear thinning and shear thickening fluids; however, the velocity profile is unaffected in the vicinity of the disk. Figures 3(a,b) show the profiles of azimuthal component $g(\eta)$ of the velocity field for different values of the power-law index $n$. The velocity approaches the asymptotic value, monotonically for each value of $n$. The velocity also decreases with each increment of $n$ due to increase in viscous forces. Further, the velocity profiles become closer as the value of the power-law index is extended beyond $n > 1$. The variation in axial component $h(\eta)$ of velocity is sketched in figures 4(a,b) for different values of the power-law index $n$. The velocity profile decreases as the value of $n$ is incremented due to the fact that the radial velocity component decreases as the value of $n$ is increased. Further, this figure also depicts that for highly shear thinning fluid the value of $h(\eta)$ fails to approach an asymptotic limit as attained by other two components $f(\eta)$ and $g(\eta)$ for large value of $\eta$.

The heat transfer aspects of the Sisko fluid over a rotating constant surface temperature disk possessing constant surface temperature for shear thinning and thickening fluids for different values of the power-law index $n$ is illustrated in figures 5(a,b). Figure 5(a) depicts that the temperature profile and thermal boundary layer reduce with incrementing value of the power-law index $n < 1$. The effect on the temperature profile diminishes when the power-law index approaches unity. Figure 5(b) reveals that the power-law index $n$ does not affect the temperature profile as strongly as for shear thinning fluid. However, a decrease in the thermal boundary layer thickness is observed and temperature profiles become closer, as the value of $n$ is incremented.

To exhibit the effects of variation in the material parameter $A$ of Sisko fluid on non-dimensional velocities $f(\eta)$, $g(\eta)$ and $h(\eta)$ profiles, we have plotted figures 6 to 8. These figures show that the velocity increases as the value of $A$ is augmented. For shear thickening Sisko fluid (panel $b$), the material parameter $A$ of Sisko fluid affects more strongly in contrast to the shear thinning fluid (panel $(a)$). Figures 9(a,b) illustrate the effect of the material parameter $A$ of Sisko fluid on temperature field. These figures show that temperature profile is not considerably affected by the variation in $A$; nevertheless, the temperature is slightly lowered when value of $A$ is increased progressively.

The Prandtl number $\Pr$ of a fluid plays a vital role in forced convective heat transfer. Figures 10(a,b) present its effect on heat transfer to Sisko fluid for shear thinning and shear thickening regimes. These figures depict that Pr affects the heat transfer process strongly by thinning the thermal boundary layer thickness. It in turn augments the heat transfer at the wall. The augmentation can be ascribed to the enhanced momentum diffusivity for larger Prandtl number. The temperature profile is slightly lower for fluids with shear thickening behavior than that of the shear thinning for same Prandtl number.

Lastly, figures 11 and 12 make a comparison amongst the velocities and temperature profiles of the Newtonian fluid $(A=0$ and $n=1)$ and the power-law fluid $(A=0$ and $n\neq1)$ with those of the Sisko fluid $(A\neq0)$. Figures 11(a-c) reveal that the velocity distribution for the Newtonian and the power-law $(n=1.2)$ fluids are close to each other, but that of Sisko fluid has a thick boundary layer. One can deduce that the material parameter $A$ of Sisko fluid has a decisive influence on the boundary layer thickness. The thermal boundary layer thickness for the Sisko and the power-law fluids is thin as compare to that of the Newtonian fluid. But thermal boundary layer thickness in not much altered by the material parameter of Sisko fluid.

Table 2 summarizes the overall trends in the radial and azimuthal velocity gradients at the disk, based on the variation in the power-law index $n$. This table shows that the velocity gradient at the disk increases, but that is not a strong function of $n$. Further, this table also depicts that axial velocity out the boundary layer heavily dependent on $n$, and decreases as the value of the power-law index $n$ is augmented due to decrease in the radial component of velocity. The effect of variation in the Prandtl number $\mathrm{Pr}$ on the temperature gradient at the disk surface is outlined in table 3. It can be noted that the temperature gradient increases with rise in Pr; however, the gradients at the disk increase $49\%$ and $55\%$ for shear thinning and shear thickening Sisko fluids, respectively.

***Table 1(a):*** A comparison of the radial velocity gradient at wall when $A = 0$ is fixed.

| Power-law index | $f'(0)$ | | |
|---|---|---|---|
| $n$ | Present result | Anderson *et al.* [14] | M&U [12] |
| 0.2 | 0.529 | 0.532 | 0.528 |
| 0.4 | 0.504 | - | 0.504 |
| 0.5 | 0.501 | 0.501 | 0.501 |
| 0.6 | 0.501 | 0.501 | 0.500 |
| 0.8 | 0.504 | 0.504 | 0.504 |
| 0.9 | 0.507 | 0.507 | 0.507 |
| 1.0 | 0.510 | 0.510 | 0.510 |
| 1.3 | 0.521 | 0.522 | 0.521 |
| 1.5 | 0.529 | 0.529 | 0.529 |
| 1.7 | 0.537 | 0.537 | - |

***Table 1(b):*** A comparison of the azimuthal velocity gradient at wall when $A = 0$ is fixed.

| *Power law index* | $-g'(0)$ | | |
|---|---|---|---|
| $n$ | Present results | Anderson *et al.*[14] | M&U [12] |
| 0.2 | 1.034 | 1.032 | 1.037 |
| 0.4 | 0.769 | - | 0.769 |
| 0.5 | 0.713 | 0.712 | 0.713 |
| 0.6 | 0.678 | 0.676 | 0.677 |
| 0.8 | 0.636 | 0.636 | 0.636 |
| 0.9 | 0.624 | 0.624 | 0.624 |
| 1.0 | 0.616 | 0.616 | 0.616 |
| 1.3 | 0.603 | 0.603 | 0.603 |
| 1.5 | 0.601 | 0.601 | 0.601 |
| 1.7 | 0.601 | 0.600 | - |

***Table 1(c):*** A comparison of the axial velocity when $A = 0$ is fixed.

| Power law index | $-h(\infty)$ | | |
| n | Present results | Anderson *et al.* [14] | M&U [12] |
|---|---|---|---|
| 0.2 | 1.544 | - | - |
| 0.4 | 1.712 | - | - |
| 0.5 | 1.592 | 1.539 | 1.513 |
| 0.6 | 1.430 | 1.364 | 1.351 |
| 0.8 | 1.089 | 1.089 | 1.052 |
| 0.9 | 0.970 | 0.969 | 0.958 |
| 1.0 | 0.884 | 0.883 | - |
| 1.3 | 0.736 | 0.735 | 0.735 |
| 1.5 | 0.678 | 0.676 | 0.678 |
| 1.7 | 0.637 | 0.633 | - |

***Table 2:*** A tabulation of the velocity gradient at wall when $A = 1.5$ is fixed.

| n | $f'(0)$ | $-g'(0)$ | $-h(\infty)$ |
|---|---|---|---|
| 0.4 | 0.29599 | 0.38973 | 2.57607 |
| 0.6 | 0.30717 | 0.38534 | 1.92475 |
| 0.8 | 0.31559 | 0.38605 | 1.54620 |
| 1.2 | 0.32895 | 0.39443 | 1.32478 |
| 1.4 | 0.33452 | 0.39995 | 1.28217 |
| 1.6 | 0.33951 | 0.40570 | 1.25597 |
| 1.8 | 0.34395 | 0.41146 | 1.23178 |

***Table 3***: A tabulation of the local Nusselt number when $A = 1.5$ fixed.

| | $-\theta'(0)$ | |
| --- | --- | --- |
| Pr | $n = 0.5$ | $n = 1.5$ |
| 3.0 | 0.57042 | 0.71071 |
| 5.0 | 0.68979 | 0.87693 |
| 7.0 | 0.77976 | 1.00293 |
| 9.0 | 0.85363 | 1.10669 |

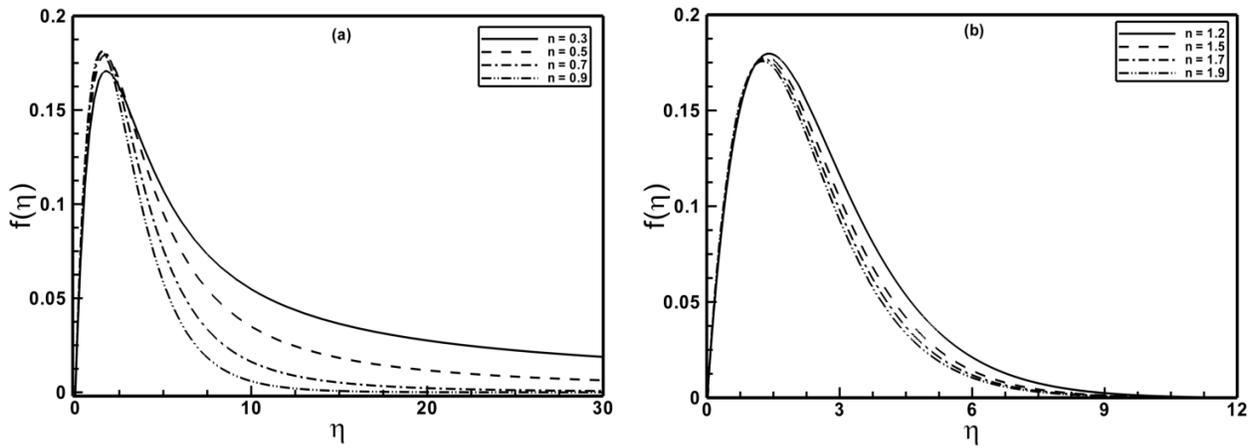

**Figure 2:** Profiles of the radial velocity component $f(\eta)$ for different values of the power-law index when $A = 1.5$ is fixed.

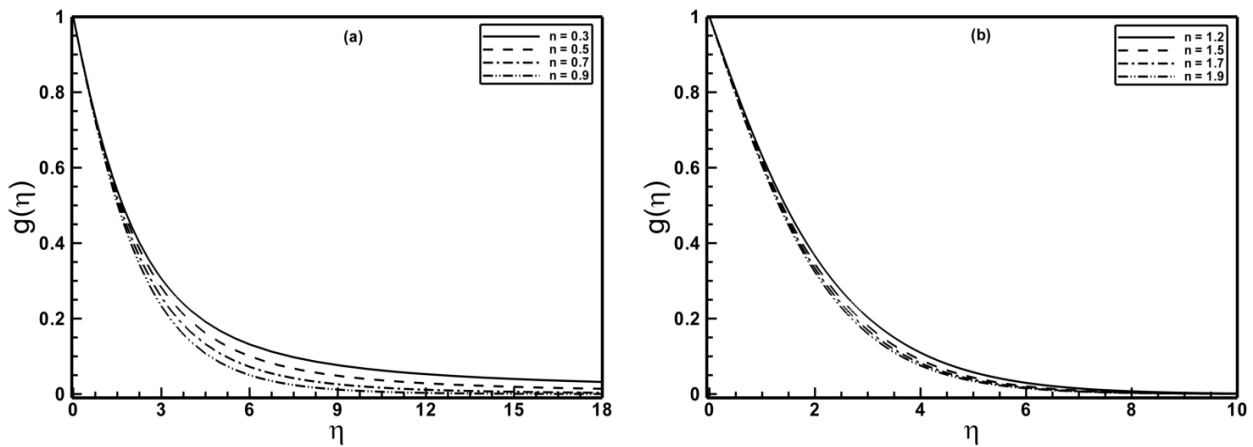

**Figure 3:** Profiles of the azimuthal velocity component $g(\eta)$ for different values of the power-law index when $A = 1.5$ is fixed.

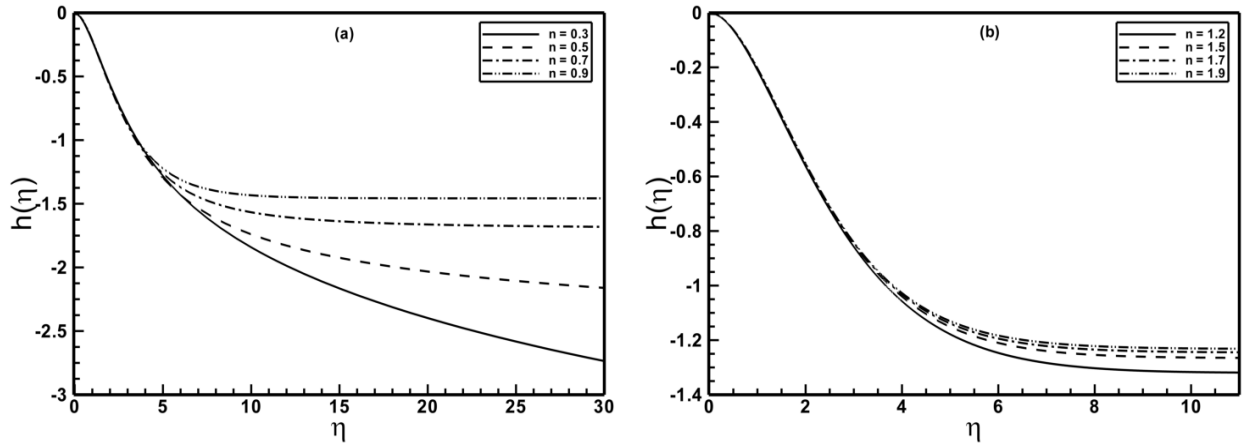

***Figure 4:*** Profiles of the axial velocity component $h(\eta)$ for different values of the power-law index when $A = 1.5$ is fixed.

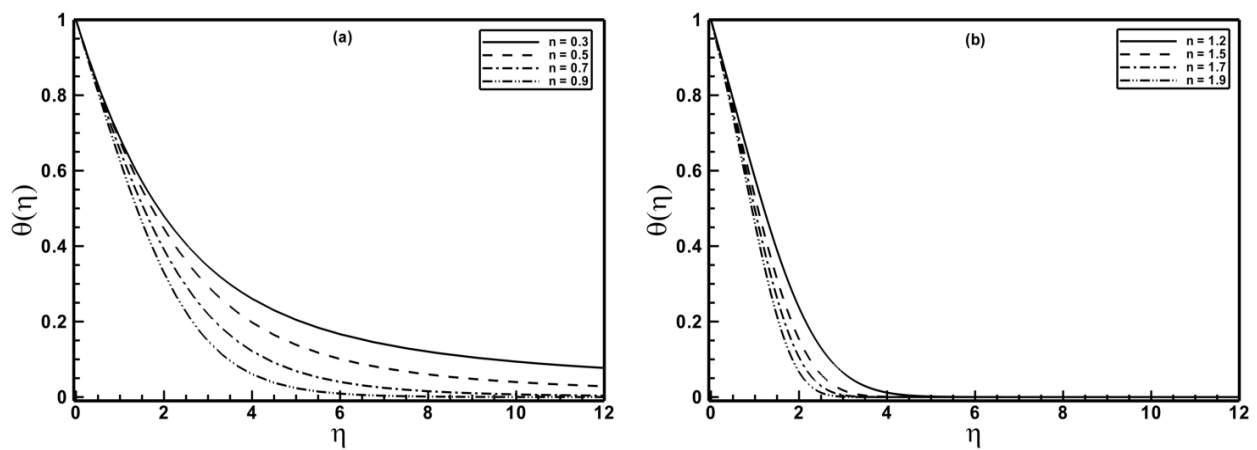

***Figure 5:*** Profiles of the temperature $\theta(\eta)$ for different values of the power-law index when $A = 1.5$ and $\text{Pr} = 1.0$ are fixed.

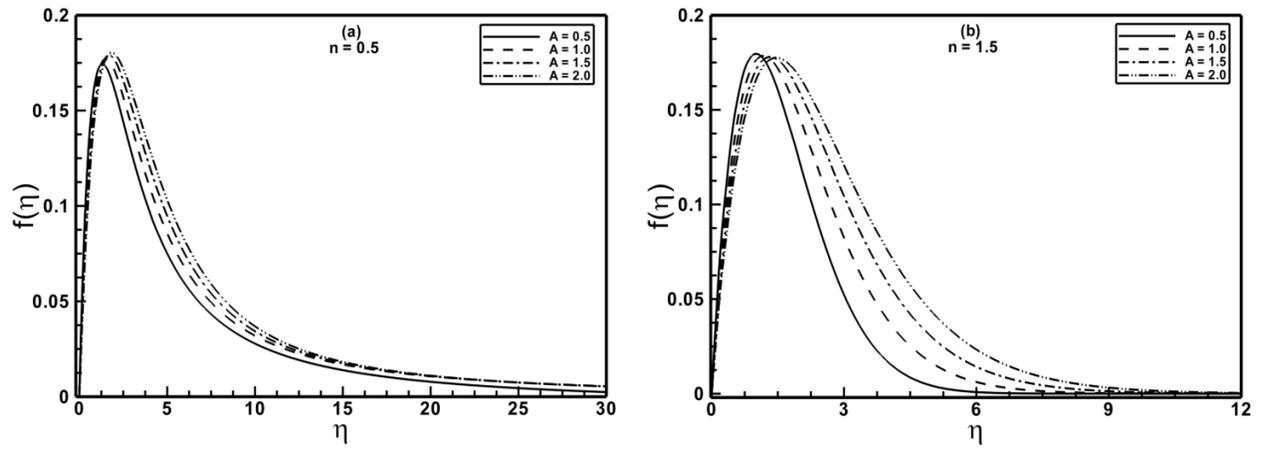

**Figure 6:** Profiles of the radial velocity component $f(\eta)$ for different values of the material parameter $A$ of Sisko fluid.

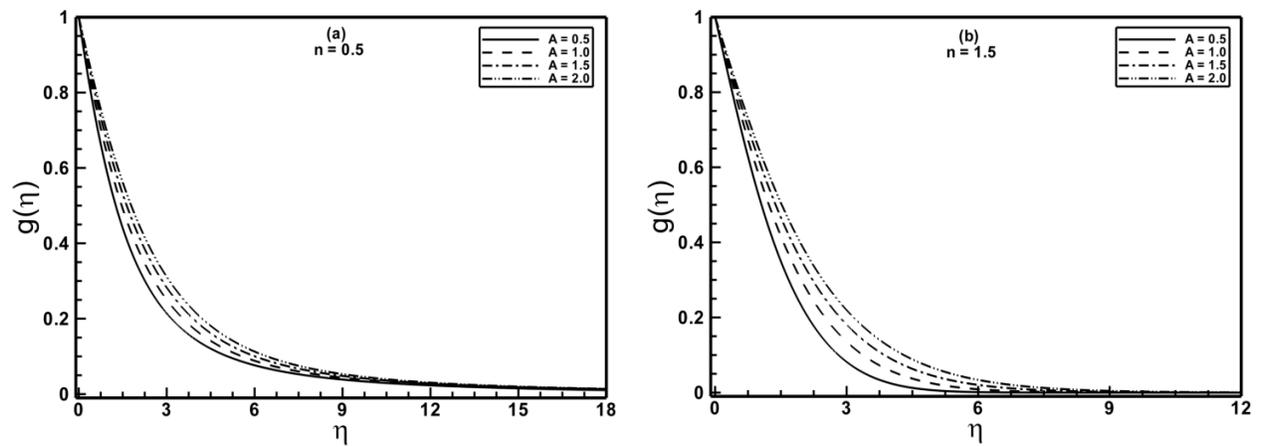

**Figure 7:** Profiles of the azimuthal velocity component $g(\eta)$ for different values of the material parameter $A$ of Sisko fluid.

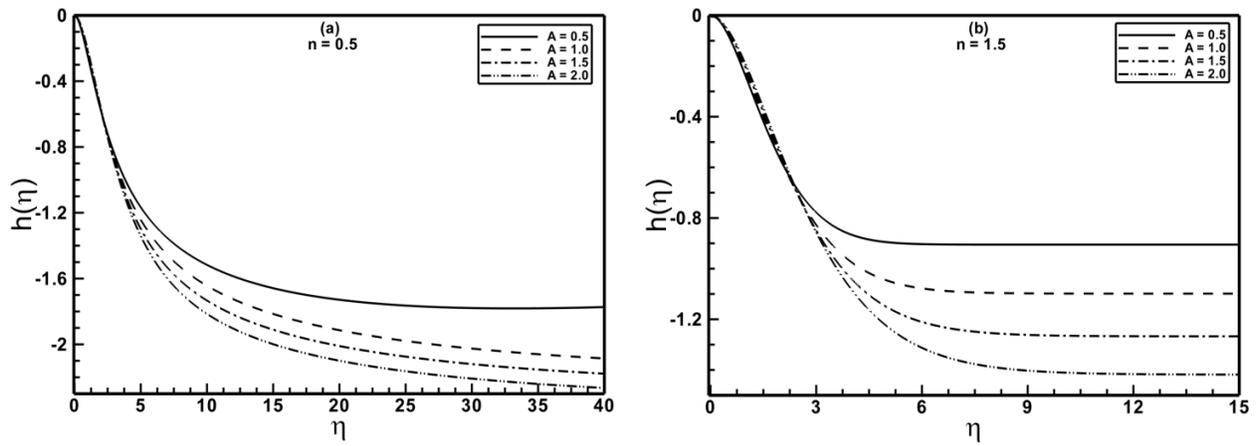

**Figure 8:** Profiles of the axial velocity component $h(\eta)$ for different values of the material parameter $A$ of Sisko fluid.

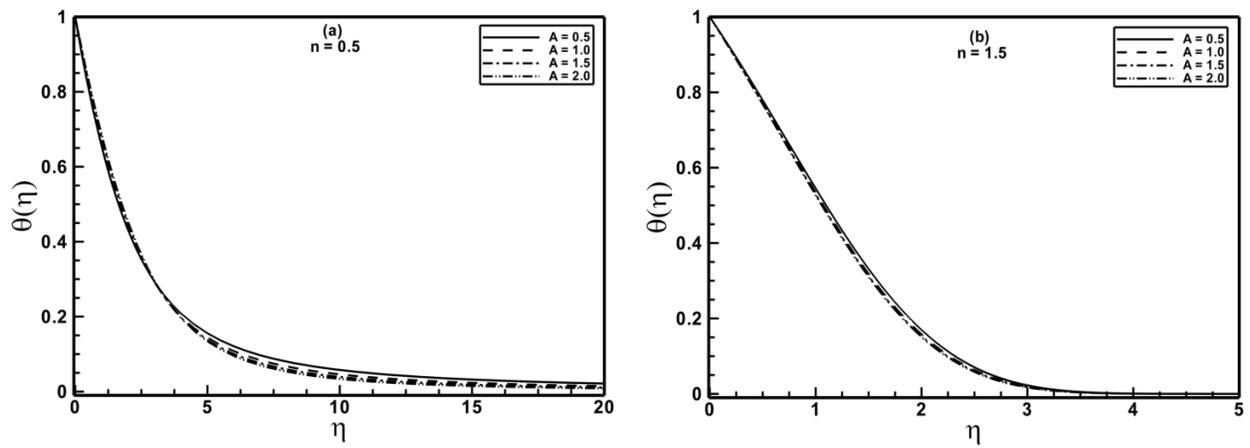

**Figure 9:** Profiles of the temperature $\theta(\eta)$ for different values of the material parameter $A$ of Sisko fluid when $\mathrm{Pr} = 1.5$ is fixed.

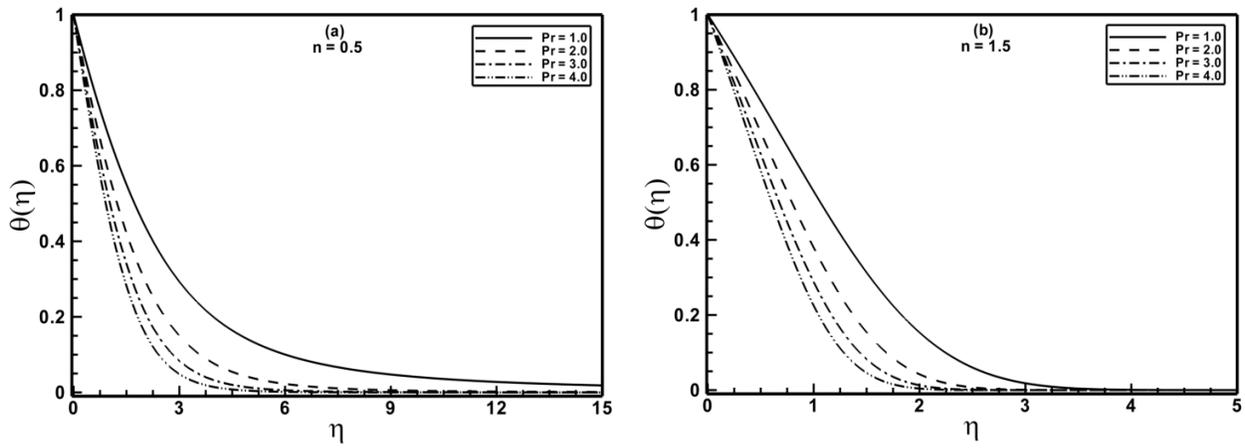

***Figure 10:*** Profiles of the temperature $\theta(\eta)$ for different values of the Prandtl number $_{Pr}$ when $A = 1.5$ is fixed.

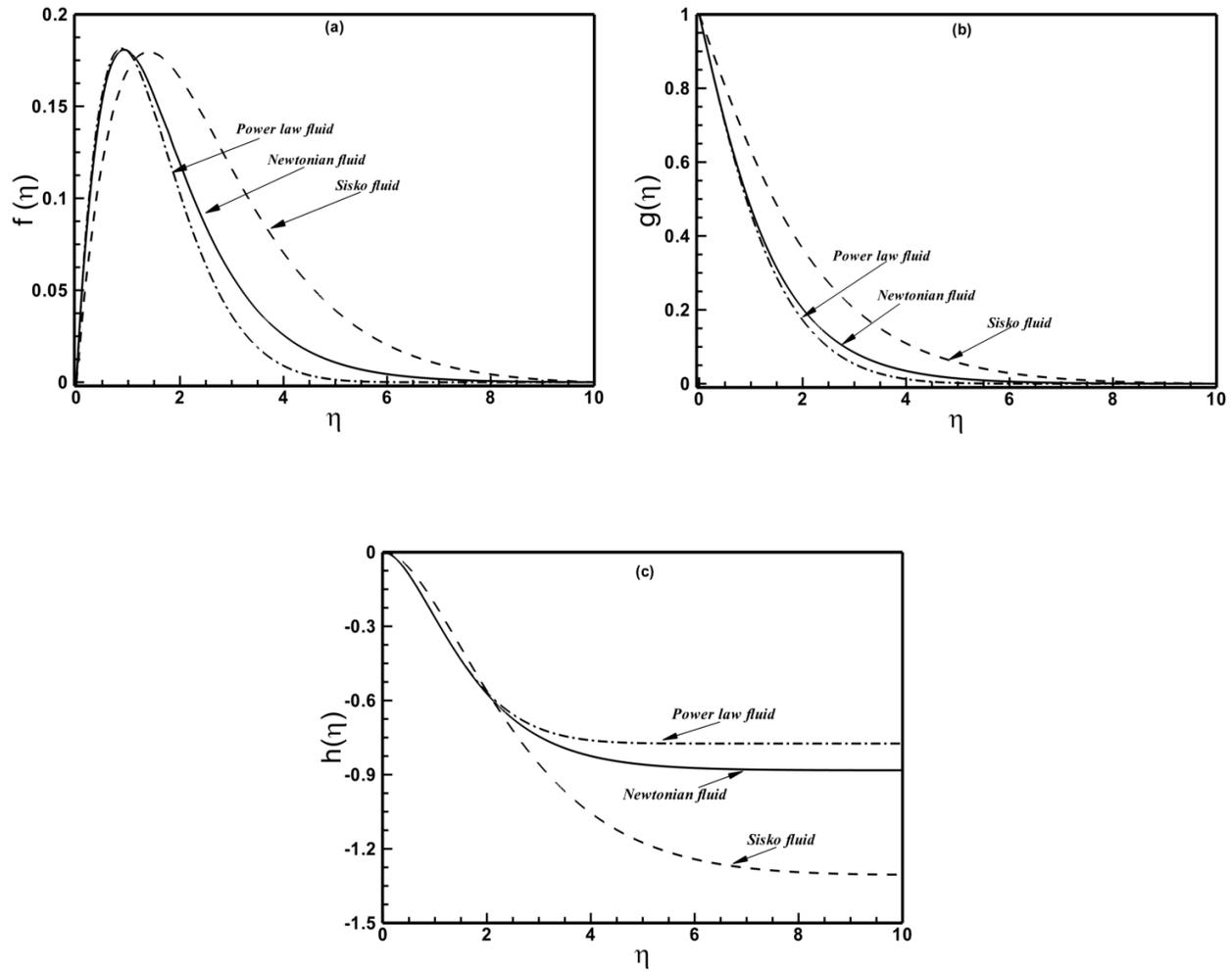

***Figure 11:*** A comparison of the radial, azimuthal and axial velocity components for different fluids.

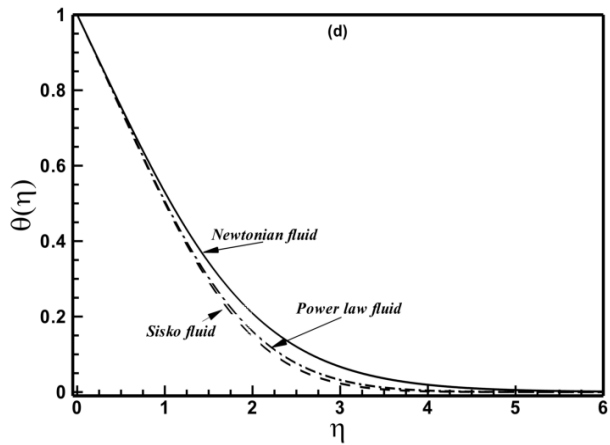

***Figure 12:*** A comparison of the temperature distribution for different fluids.

## 6. Conclusions

The steady three-dimensional flow and heat transfer characteristics within boundary layer of Sisko fluid over an isothermal infinite rotating disk has been studied numerically. The effects of the power-law index, the material parameter and the Prandtl number on the velocity and temperature profiles were studied. Moreover, velocity and temperature profiles of the power-law, Newtonian and Sisko fluids have also been compared. Our computations have showed that:

- The radial, azimuthal and axial velocity components have decreased as the power-law index was augmented and the velocity profiles were more sensitive for shear thinning Sisko fluid.

- The characteristic peak in the radial velocity profiles, caused due to radial outward flow caused due to centrifugal force, has increasing trend with raising the power-law index for shear thinning Sisko fluid.

- A thinning of the thermal boundary layer was seen for increasing power-law index, and hence better heat transfer at the wall.

- An increase in Prandtl number resulted in thinner thermal boundary layer and the thinning of boundary layer was more prominent for shear thickening Sisko fluid for same range of Prandtl number.

- The thickening of boundary layer was resulted in by following increase in the value of material parameter of Sisko fluid and increase in boundary layer thickness is more prominent for shear thickening Sisko fluid.

- The Sisko fluid has thickest boundary layer and thinnest thermal boundary layer when compared with that Newtonian and power-law fluid.